\documentclass[a4paper,12pt]{article}
\usepackage{amsmath}
\usepackage{amssymb}
\usepackage{amsfonts}
\usepackage{amscd}
\usepackage{bbm} 
\usepackage{fleqn} 
\usepackage[footnotesize]{caption}
\usepackage{graphicx} 
\usepackage{braket} 
\usepackage{mathrsfs}
\usepackage[center,footnotesize,hang]{subfigure}
\usepackage{BetterCite}
\usepackage{accents}

\newcommand{\CenterFmg}[1]{\vcenter{\hbox{\includegraphics{#1.eps}}}}

\newcommand{\CenterEps}[2][1]{\vcenter{\hbox{\includegraphics[scale=#1]{#2.eps}}}}

\def\<{\left\langle}
\def\>{\right\rangle}

\def\ChargeC{\mathrm{C}}

\newcommand{\RaiseBrace}[1]{\raise3pt\hbox{$\displaystyle#1$}}

\DeclareMathOperator{\Tr}{Tr}

\allowdisplaybreaks[1]

\begin{document}

\begin{titlepage}

\ \vspace*{-15mm}
\begin{flushright}
TUM-HEP-491/02\\
DESY 02-208
\end{flushright}
\vspace*{5mm}

\begin{center}
{\Large\sffamily\bfseries 
 Dynamical Electroweak Symmetry Breaking \\[2mm] 
 by a Neutrino Condensate
}\\[12mm]
Stefan Antusch\footnote{E-mail: \texttt{santusch@ph.tum.de}},
J\"{o}rn Kersten\footnote{E-mail: \texttt{jkersten@ph.tum.de}},
Manfred Lindner\footnote{E-mail: \texttt{lindner@ph.tum.de}}
\\
{\small\it
Physik-Department T30, 
Technische Universit\"{a}t M\"{u}nchen\\ 
James-Franck-Stra{\ss}e,
85748 Garching, Germany
}
\\[3mm]
Michael Ratz\footnote{E-mail: \texttt{mratz@mail.desy.de}}\\
{\small\it
Deutsches Elektronensynchrotron DESY\\ 
22603 Hamburg, Germany
}
\end{center}
\vspace*{2.0cm}

\begin{abstract}
\noindent
We show that the electroweak symmetry can be broken in a
natural and phenomenologically acceptable way by a neutrino 
condensate. Therefore, we assume as particle content only 
the chiral fermions and gauge bosons of the Standard Model 
and in addition right-handed neutrinos. A fundamental Higgs 
field is absent. We assume instead
that new interactions exist that can effectively be described 
as four-fermion interactions and that can 
become critical in the neutrino sector. We discuss in detail 
the coupled Dirac-Majorana gap equations which lead to a 
neutrino condensate, electroweak symmetry breaking and via 
the dynamical see-saw mechanism to small neutrino masses.
We show that the effective 
Lagrangian is that of the Standard Model with massive neutrinos 
and with a composite Higgs particle. The mass predictions are 
consistent with data. 
\end{abstract}

\end{titlepage}

\newpage
\setcounter{footnote}{0}

\section{Introduction}
The generalization of renormalizable relativistic gauge theories 
to the Standard Model (SM) was very successful and has been confirmed
experimentally in an impressive way, including detailed tests of
radiative corrections.\footnote{Recent measurements of small deviations 
of $g\!-\!2$, $A_\mathrm{FB}$ and $\sin^2\Theta_W$ in the neutrino 
sector may be the first signs of physics beyond 
the SM \cite{Langacker:2002sy}.} However, it is important to keep in mind that 
the mechanism of electroweak (EW) symmetry breaking is still mostly 
untested. The postulated Higgs particle has so far not been observed 
and there is only indirect 
evidence from quantum corrections that a SM Higgs boson 
should be lighter than about $200\,\mathrm{GeV}$ \cite{Higgsbound}. 
The Higgs sector has 
furthermore well-known theoretical problems, especially the gauge 
hierarchy problem, which strongly suggest that new physics exists 
which is connected to the mechanism of EW symmetry breaking. 
Whatever the correct symmetry breaking mechanism is, it must satisfy 
by now a number of stringent direct and indirect constraints. Given 
the success of the SM it is, however, immediately clear how an 
alternative symmetry breaking scenario can be consistent with 
data. In the limit where new physics decouples, it just has 
to reproduce effectively the SM Higgs sector with a light Higgs 
particle \cite{Lindner:1997sd}. If the model has such a decoupling limit, 
as in the case discussed in this paper, then deviations from the SM can 
be understood as a departure from the decoupling limit.

Motivated by the evidence for neutrino masses, we discuss the possibility 
that the EW symmetry is broken dynamically by a neutrino condensate. 
This would normally lead to neutrino masses of the order of the 
symmetry breaking scale, i.e.\ \(\mathscr{O}(200\,\mathrm{GeV})\). 
Neutrinos may, however, possess both Dirac and Majorana mass terms 
and the dynamical generation of large Dirac mass terms leads via the 
see-saw mechanism \cite{Gell-Mann:1980vs,Yanagida:1980,Mohapatra:1980ia} 
to small, phenomenologically acceptable neutrino masses. A composite 
Higgs particle will emerge that is not affected by the see-saw mechanism, 
i.e.\ it will have a mass of the order of the EW symmetry breaking scale. 
The low-energy effective Lagrangian in the decoupling limit is therefore
the SM, with a composite Higgs instead of a fundamental scalar. 

A heavy Dirac neutrino mass is similar to the heavy top mass
of the order of the EW scale, which gave rise to 
speculations that top condensation might be responsible for EW 
symmetry breaking \cite{Bardeen:1990ds,Lindner:1993ah}. However, top
condensation is not viable in its simplest version, since it predicts
too large top and Higgs masses. Different non-minimal models are in
principle viable \cite{Luty:1990bg,Carena:1992ky,Akhmedov:1996vm,Akhmedov:1996ip},
and the possibility that third-generation neutrinos contribute to 
top condensation was studied by Martin \cite{Martin:1991xw}. The 
condensation of a full fourth generation (including neutrinos of the fourth 
generation) \cite{Hill:1991ge} was also discussed in this context.
We study the case where only neutrinos are responsible for
the dynamical breakdown of the EW symmetry. We briefly 
discuss a mixed case as well where combined neutrino and top
condensation leads to an effective two-Higgs scenario with a
leptonic and a hadronic Higgs particle.

The paper is organized as follows: In the next section we discuss the
condensation of neutrinos, i.e.\ we study the relevant system of coupled
gap equations in combination with the see-saw mechanism in the proper
mass eigenstate basis. Afterwards, we solve the gap numerically.  The
following section contains the phenomenology and the predictions arising
from the renormalization group improved compositeness conditions.
Section~\ref{sec:3neutrinos} contains a short discussion of the option
that all three generations of neutrinos condense simultaneously and in
section~\ref{sec:2higgs} we outline briefly the possibility of a
combined neutrino-top condensation scenario.

\section{Neutrino Condensation} \label{sec:NeutrinoCondensation}

We assume as mentioned that some physics exists at high energies
which yields an effective four-fermion picture
similar to weak interactions at low energies. However, contrary to weak 
interactions we assume that certain four-fermion couplings
become strong enough to trigger the formation of condensates, thus
giving masses to some of the fermions via gap equations.  
The remaining fermions could e.g.\ obtain masses from further
four-fermion couplings, which only subdominantly contribute to the gap.
For top condensation it was shown how this can be justified 
in the context of broken renormalizable gauge theories at high energies,
for example in the framework of so-called top-color models 
\cite{Hill:1991at} or U(1) models \cite{Lindner:1992bs}.
In this spirit, we consider the particle content of the SM extended by 
three right-handed neutrinos but without a fundamental Higgs field. 
Instead of the SM Higgs field we assume four-fermion couplings involving
the lepton doublets and the right-handed neutrinos. In addition, since 
the right-handed neutrinos are singlets under the SM gauge group and 
since there is no protective symmetry, we assume them to have huge 
Majorana masses. 

In order to show the essential aspects of such a scenario,
we consider first the case where only one of the four-fermion couplings
drives the condensation, while the others vanish. Thus, the four-fermion 
Lagrangian is 
\begin{eqnarray}\label{eq:4FNeutrino}
  \mathscr{L}_\mathrm{4f} = G^{(\nu)} \, (\overline{ \ell_{\mathrm{L}}} 
  \nu_{\mathrm{R}})
  (\overline{ \nu_{\mathrm{R}}} \ell_{\mathrm{L}}) 
\;,
\end{eqnarray}
where we have omitted the \(\mathrm{SU}(2)\) indices and where
$\ell_\mathrm{L}$, $\nu_\mathrm{R}$ stand for the relevant 
neutrino degrees of freedom. 
Moreover, we assume the Majorana mass matrix to be diagonal so that
the condensing pair of neutrinos can be studied independently.
Therefore, we need to consider only one Majorana mass term,
\begin{equation}\label{eq:DirectMajoranaMass}
 -\mathscr{L}_\mathrm{M}=\frac{1}{2}M \, \overline{\nu_{\mathrm{R}}}\nu^\ChargeC_{\mathrm{R}}
 +\text{h.c.}\; .
\label{eq:MajoranaMass}
\end{equation}
We will see that this describes the most interesting features of neutrino
condensation. More general scenarios will be discussed briefly in 
sections~\ref{sec:3neutrinos}~and~\ref{sec:2higgs}.

\begin{figure}[ht]
\begin{center}
 \begin{eqnarray*}
  \CenterEps[0.7]{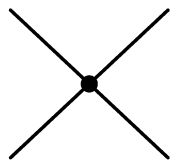}
  +
  \CenterEps[0.7]{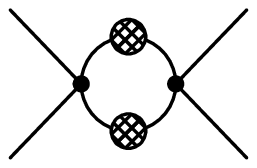}
  +
  \CenterEps[0.7]{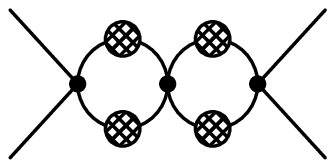}
  + \dots
  & = &
  \CenterEps[0.7]{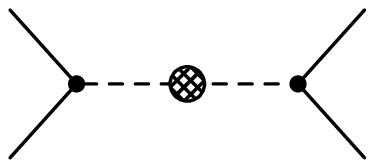}
 \end{eqnarray*}
\end{center}
\label{F1}
\caption{\label{fig:BubbleSumHiggsExchange} 
The exchange of a virtual composite Higgs scalar can be seen as a sum over 
all loop contributions involving the four-fermion vertex in the so-called 
bubble sum approximation.  Hatched blobs denote full propagators.
}
\end{figure}

\begin{figure}[ht*]
\begin{center}
\setlength{\mathindent}{3mm}
\begin{eqnarray*}
\lefteqn{
        \CenterEps[0.7]{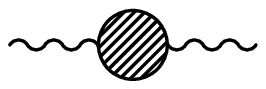} =
}
\\
& = &
        \CenterEps[0.7]{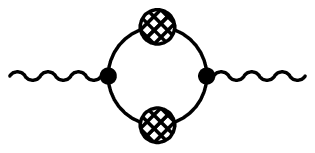} +
        \CenterEps[0.7]{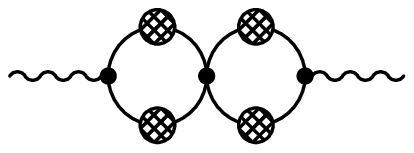} +
        \CenterEps[0.7]{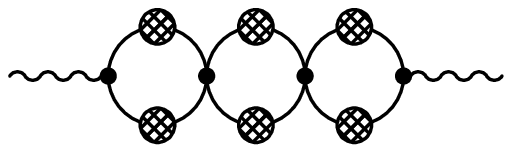} + \dots
\\[2mm]
& = &
        \CenterEps[0.7]{DynSB89} +
        \CenterEps[0.7]{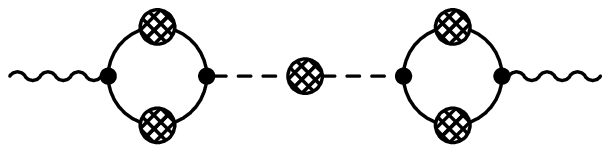}
\end{eqnarray*}
\end{center}
\label{F2}
\caption{Dynamical generation of gauge boson masses.  The bubble sum in 
 the second line is expressed in terms of the composite Higgs propagator
 from fig.~\ref{fig:BubbleSumHiggsExchange} in the third line.
 The shaded blob on the left side is the OPI 2-point vertex function. 
 \label{fig:BuggleSumGaugeBosonMass}
}
\end{figure}

The question whether a non-perturbative solution for the ground state exists 
in the presence of the huge Majorana mass, i.e.\ if the gap equation 
has a non-trivial solution, will be studied in section~\ref{sec:gap}. 
If the gap equation produces a fermion condensate which is a doublet under 
\(\mathrm{SU}(2)_\mathrm{L}\) and which carries a suitable 
$\mathrm{U}(1)_\mathrm{Y}$ charge, then it is immediately 
clear that the \(\mathrm{SU}(2)_\mathrm{L} \otimes \mathrm{U}(1)_\mathrm{Y}\) 
gauge symmetry is broken. For the chosen NJL-like interaction a composite
Higgs particle emerges. This is visualized in fig.~\ref{fig:BubbleSumHiggsExchange}, 
where a massive scalar pole and three massless 
Goldstone bosons are produced by the summation of a certain class of diagrams 
with dynamical fermion propagators. The Higgs mechanism and the ``eating'' of
the Goldstone bosons is illustrated in fig.~\ref{fig:BuggleSumGaugeBosonMass}.
For more details see e.g.~\cite{Lindner:1993ah}.

We analyze now the gap equations of our model with an explicit Majorana 
mass term for \(\nu_\mathrm{R}\). If a non-trivial solution produces dynamically a 
large Dirac mass term, it breaks the EW
symmetry. Then the presence of the huge singlet Majorana 
mass will lead to a dynamical see-saw mechanism with small neutrino masses. 
A computation of the gap equation in the basis of mass eigenstates must 
therefore include in a self-consistent way the possibility of a dynamically 
generated Dirac mass term, 
\begin{eqnarray}
-\mathscr{L}_\mathrm{D} = D \, \overline{
\nu_{\mathrm{L}}}\nu_{\mathrm{R}} +\text{h.c.}\;.
\end{eqnarray}

\subsection{Mass Eigenstates and Eigenvalues}

For any value of \(D\), the mass eigenstates are two Majorana fermions, given by
\begin{eqnarray}\label{eq:MassEigMajDirCase}
\left(\begin{array}{c}\lambda\\ \rho\end{array}\right)
 =  U\cdot\left(\begin{array}{c}\lambda'\\ \rho'\end{array}\right)
\end{eqnarray}
with $\lambda':= \nu_{\mathrm{L}}+\nu^\ChargeC_{\mathrm{L}}$ 
and $\rho':=\nu_{\mathrm{R}}+
\nu^\ChargeC_{\mathrm{R}}$ 
and the orthogonal matrix
\begin{equation}\label{eq:ParametrizationOfUforRealDandM}
 U=
 \left(\begin{array}{cc}\cos\varphi & \sin\varphi\\
 -\sin\varphi & \cos\varphi\end{array}\right)
 =:\left(\begin{array}{cc}
        c & s \\ -s & c
 \end{array}\right)\;.
\end{equation}
The corresponding mass eigenvalues
\begin{subequations} \label{eq:MassEigenvalues}
\begin{eqnarray}
m_\lambda&=&\frac{1}{2}\left(M- \sqrt{4D^2+M^2}\right)\;, \\
m_\rho&=&\frac{1}{2}\left(M+\sqrt{4D^2+M^2}\right)
\end{eqnarray}
\end{subequations}
are related to \(\varphi\) by
\begin{equation} \label{eq:DefPhi}
 \varphi = \arctan \sqrt{-\frac{m_\lambda}{m_\rho}}\;.
\end{equation}
For convenience, we rewrite the singlet Majorana mass term 
\eqref{eq:DirectMajoranaMass} as well as the neutrino
part of the four-fermion interaction \eqref{eq:4FNeutrino} 
in terms of Majorana fermions, 
\begin{eqnarray}
 -\mathscr{L}_\mathrm{M}
 & = &
 \frac{1}{2} M\,\overline{\nu_\mathrm{R}}\nu_\mathrm{R}^\ChargeC+\text{h.c.}
 \:\:=\:\:
 \frac{1}{2} M\,\overline{\rho'}\,P_\mathrm{R}\,{\rho'}^\ChargeC+\text{h.c.}
 \;,
 \label{eq:DirectMassTerm}
\\
\mathscr{L}_{4\nu} &=& G^{(\nu)} \, (\overline{ \nu_{\mathrm{L}}}  \nu_{\mathrm{R}})
\,  (\overline{ \nu_{\mathrm{R}}} \nu_{\mathrm{L}}) \:\:=\:\:
G^{(\nu)} \, (\overline{\lambda'}\, P_\mathrm{R}\, \rho' )\,
(\overline{\rho'}\,P_\mathrm{L}\, \lambda') \;.
\end{eqnarray}
The Feynman rules for the interactions of the mass eigenstates are
derived from these Lagrangians by inserting the relations of 
eq.~\eqref{eq:MassEigMajDirCase}.

\subsection{The Coupled Gap Equations} \label{sec:gap}
\begin{figure}[h*]
\begin{center}
$
  \CenterFmg{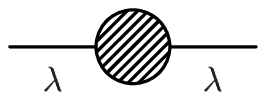} 
  =
  \begin{array}{c}
  \\[-2mm]
  \CenterFmg{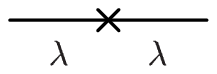} 
  \end{array}
  +
  \CenterFmg{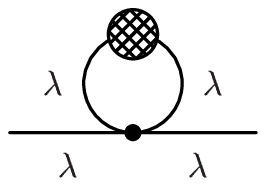} 
  +
  \CenterFmg{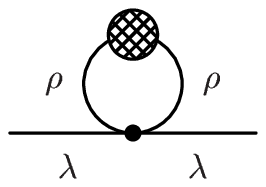}
$\\[3mm]

$
  \CenterFmg{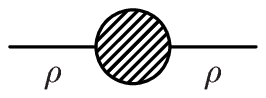} 
  =
  \begin{array}{c}
  \\[-2mm]
  \CenterFmg{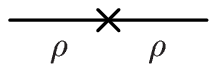} 
  \end{array}
  +
  \CenterFmg{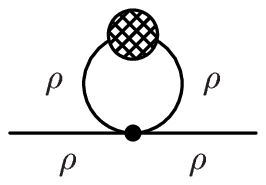} 
  +
  \CenterFmg{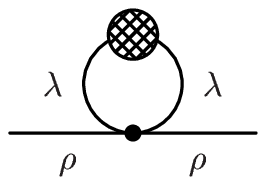}
$
\end{center}
\caption{\label{fig:gap} Gap equations for $m_\lambda$, the mass eigenvalue of the
light neutrino, and $m_\rho$, the mass eigenvalue of the heavy neutrino.
The explicit Majorana mass from eq.~\eqref{eq:DirectMassTerm}
is indicated by a cross. The shaded blobs on the left side are the OPI 2-point 
vertex functions, whereas the hatched blobs on the right side are full propagators.  
}
\end{figure}

The gap equations for the masses $m_\lambda$ and $m_\rho$, corresponding to
fig.~\ref{fig:gap}, are
\begin{subequations} \label{eq:GapEquations}
\begin{eqnarray}
\label{eq:GapForMLambda}
 m_\lambda  & = &  2 \,G^{(\nu)}\,c^2\,s^2\,
 \left[
        m_\lambda\,I_\mathrm{gap}(m_\lambda)
        -m_\rho\,I_\mathrm{gap}(m_\rho)
 \right]
 +s^2\,M\;,\\
 m_\rho  & = & 2 \,G^{(\nu)}
 \,c^2\,s^2\,\left[
        m_\rho\,I_\mathrm{gap}(m_\rho)
        -m_\lambda\,I_\mathrm{gap}(m_\lambda)
 \right]
 +c^2\,M \;,
\end{eqnarray}
\end{subequations}
where we have introduced
\begin{equation}
 \tfrac{1}{2}\,I_\mathrm{gap}(m)
 :=
 - \frac{\Lambda^2}{16\pi^2}
 \left[1-\frac{m^2}{\Lambda^2} \ln \left(\frac{\Lambda^2}{m^2}+1\right)\right]
 \;.
\end{equation}
\(\Lambda\) is the condensation scale, which acts as a cutoff.
As $m_\lambda + m_\rho = M$ due to eq.~\eqref{eq:MassEigenvalues}, the
gap equations are linearly dependent, so that it is sufficient to solve
one of them.
Note that since the mass eigenvalues are given by $m_\lambda$ and $m_\rho$, 
non-trivial solutions also imply a dynamically generated Dirac mass $D$.
Such solutions are found to exist indeed, 
as will be shown in section~\ref{sec:NumGap}.
Remarkably, in this kind of gap equations, the heavy Majorana degree of
freedom plays an essential role and does not decouple.

\begin{figure}[ht*]
\begin{center}
$
 \CenterFmg{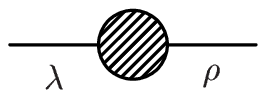}
 =
  \begin{array}{c}
  \\[-2mm]
 \CenterFmg{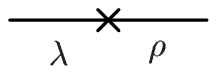}
  \end{array}
 +
 \CenterFmg{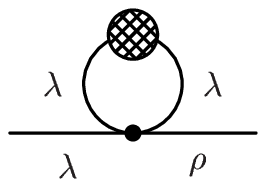}
 +
 \CenterFmg{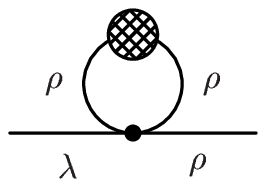}
$
\end{center}
\caption{\label{fig:consistency} Self-consistency check for the off-diagonal
elements of the gap equations in the mass eigenstate basis. We find that the 
right-hand side vanishes as required.}
\end{figure}

To check the self-consistency of our calculation, we consider the gap equation 
for a bilinear term that contains the fields $\lambda$ and $\rho$ as shown in
fig.~\ref{fig:consistency}. As we are working in the mass eigenstate
basis, this term, which corresponds to an off-diagonal entry
$m_{\lambda\rho}$ in the mass matrix, has to vanish identically.
We obtain
\begin{eqnarray}
m_{\lambda\rho}
& \! = &
 G^{(\nu)}\,
 \left[
        \cot\varphi-\tan\varphi
 \right]\,c^2\,s^2\,
 \left[
        m_\lambda\,I_\mathrm{gap}(m_\lambda)-m_\rho\,I_\mathrm{gap}(m_\rho)
 \right]
 +c\,s\,M
\nonumber\\
& \! = &
 \tfrac{1}{2}\,\cot\varphi\,\left[m_\lambda-s^2\,M\right]
 +\tfrac{1}{2}\,\tan\varphi\,\left[m_\rho-c^2\,M\right]
 +c\,s\,M
 \;.
\end{eqnarray}
Using the gap equations \eqref{eq:GapEquations} as well as the
relations \eqref{eq:MassEigenvalues} and \eqref{eq:DefPhi} for the mass
eigenvalues, it follows that $m_{\lambda\rho}$ vanishes as required.
This confirms that our solution is self-consistent.

\subsection{Numerical Solution of the Gap Equation}
\label{sec:NumGap}
\begin{figure}[ht*]
\vspace*{3mm}
        \begin{center}
                $\CenterEps[1]{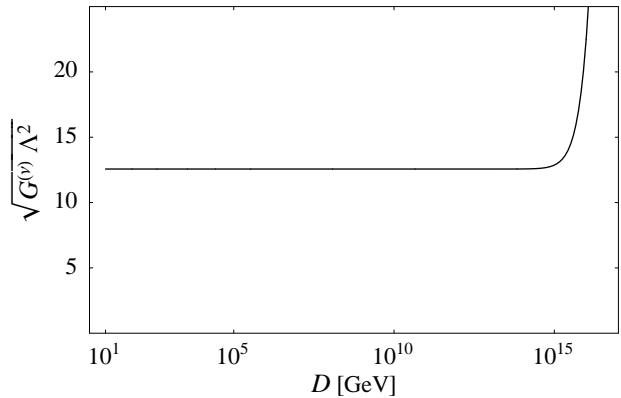}$
        \end{center}
\vspace*{-2mm}
\caption{\label{fig:MajGapNumSol} Characteristic numerical solution 
of the gap equation for the four-fermion coupling $G^{(\nu)}$ and the
dynamically generated Dirac mass $D$ with a Majorana mass
$M=10^{14}\,\mathrm{GeV}$ and a condensation scale $\Lambda=10^{16}\,\mathrm{GeV}$. 
}
\end{figure}

The gap equation \eqref{eq:GapForMLambda} for $m_\lambda$ can be considered as an equation for
$G^{(\nu)}$ and $D$, if fixed values are assigned to $M$ and $\Lambda$. 
For $M=10^{14}\,\mathrm{GeV}$ and $\Lambda=10^{16}\,\mathrm{GeV}$, 
the solution is shown in fig.~\ref{fig:MajGapNumSol}.
Instead of $G^{(\nu)}$, we plot the dimensionless coupling constant 
$g=\sqrt{G^{(\nu)}\Lambda^2}$. 
We find non-trivial solutions for $D$, if the coupling $g$ is larger than a 
critical value. This result is quite similar to top condensation \cite{Bardeen:1990ds}, 
even though the right-handed neutrino has a large Majorana mass.
Note, however, that in order to obtain correct solutions, the exact relations
\eqref{eq:MassEigenvalues} for the mass eigenvalues have to be used
rather than an expansion in powers of $\frac{D}{M}$.

In order to obtain a Dirac mass of the order of the EW scale,
fine-tuning is required in the bubble sum approximation, 
as can be seen from the extremely small slope 
of the graph in fig.~\ref{fig:MajGapNumSol}.
Even if the same fine-tuning is present in the exact gap equation, 
loop corrections, which destabilize the hierarchy in the
usual perturbative framework of the SM, do not pose an additional problem here. 
In other words, the dynamical scenario under consideration cannot explain why 
the hierarchy is large, but it does explain why it remains large. In that 
sense the hierarchy is stable and hence the model might even be considered to 
be on equal footing with the solution of the hierarchy problem by 
supersymmetry.

\section{Phenomenology} 
\label{sec:Phenomenology}
\subsection{Effective Low-Energy Theory} \label{sec:LagUebergang}

The above results for $m_\rho$ and $m_\lambda$ were obtained in
the bubble approximation.  In order to obtain more reliable low-energy
results, the renormalization group (RG) running of the 
effective theory has to be taken into account. 
Following the procedure used in \cite{Bardeen:1990ds}, we show first 
that the low-energy effective theory of the model with four-fermion 
interactions is just the SM with right-handed neutrinos.\footnote{We demonstrate this for the 
simplified case of section~\ref{sec:NeutrinoCondensation}, where all 
non-critical four-fermion couplings vanish. By including such couplings, 
the missing Yukawa interactions and masses can easily be incorporated.}
Therefore, we look at the original Lagrangian at the condensation scale $\Lambda$,
\begin{equation} \label{eq:4FLag1H}
        \mathscr{L} = 
        \mathscr{L}_\mathrm{kin} +
        \mathscr{L}_\mathrm{M} +
        G^{(\nu)} (\overline{\ell_\mathrm{L}} \nu_\mathrm{R})\,
         (\overline{\nu_\mathrm{R}} \ell_\mathrm{L}) \;,
\end{equation}
where $\mathscr{L}_\mathrm{kin}$ contains kinetic terms
for the gauge bosons, the usual SM fer\-mi\-ons
and the right-handed neutrinos.
There is no Higgs field present.
We rewrite \(\mathscr{L}\) in terms of a static (non-propagating) scalar auxiliary field $\phi$,
\begin{equation} \label{eq:AuxFieldLag1H}
        \mathscr{L} = 
        \mathscr{L}_\mathrm{kin} +
        \mathscr{L}_\mathrm{M} -
        \left( 
         \overline{\ell_\mathrm{L}} \phi \nu_\mathrm{R} + \text{h.c.}
        \right) -
        \frac{1}{G^{(\nu)}} \phi^\dagger \phi \;.
\end{equation}
This Lagrangian is equivalent to the Lagrangian in eq.~\eqref{eq:4FLag1H}, which
can be seen by exploiting the equations of motion for the auxiliary field,
\begin{equation}
        \phi = -G^{(\nu)} \, \overline{\nu_\mathrm{R}} \ell_\mathrm{L} \;.
\label{eq:hence}
\end{equation}
The same result can be found by integrating out the auxiliary field in the 
path integral formalism.

At scales below $\Lambda$, the dynamics of the theory will induce all
renormalizable and gauge invariant terms that are allowed by symmetries,
including a kinetic term and a quartic self-interaction for $\phi$.
Thus, the Lagrangian becomes
\begin{eqnarray} \label{eq:1HDMLagUnnorm}
        \mathscr{L} &=& 
        \mathscr{L}_\mathrm{kin} +
        \mathscr{L}_\mathrm{M} +
        Z\, \bigl( D_\mu \phi \bigr)^\dagger\,\bigl( D^\mu \phi \bigr) -
        \left( 
         \overline{\ell_\mathrm{L}} \phi \nu_\mathrm{R} + \text{h.c.}
        \right) +
\nonumber\\*
&& {}+
        \widetilde m^2 \, \phi^\dagger \phi -
        \frac{\widetilde\lambda}{4} \left( \phi^\dagger \phi \right)^2 \;.
\end{eqnarray}
Note that $Z$, $\widetilde m$ and $\widetilde\lambda$ are running
quantities, even though their dependence on the energy scale $\mu$ is
not written explicitly. For \(\mu\rightarrow\Lambda\) the Lagrangian 
\eqref{eq:1HDMLagUnnorm} has to become identical to the one of
eq.~\eqref{eq:AuxFieldLag1H}, which leads to the following boundary 
(compositeness) conditions for the RG evolution:
\begin{subequations} \label{eq:CompositenessUnnorm}
\begin{eqnarray}
        Z &\xrightarrow{\mu\to\Lambda}& 0 \;,
\\
        \widetilde m^2 &\xrightarrow{\mu\to\Lambda}& -\frac{1}{G^{(\nu)}} \;,
\\
        \widetilde\lambda &\xrightarrow{\mu\to\Lambda}& 0 \;.
\end{eqnarray}
\end{subequations}
Eq.~\eqref{eq:1HDMLagUnnorm} is already very similar to the SM Lagrangian. 
The auxiliary field has acquired a kinetic term, i.e.\ it has become a 
propagating composite Higgs field, but its Lagrangian is not yet written
in the usual normalization where \(Z\equiv 1\). To fix this we perform the
rescaling 
\begin{equation} \label{eq:HiggsRescaling}
        \phi \longrightarrow
        \frac{1}{\sqrt{Z}} \phi \equiv y_\nu\, \phi \;,
\end{equation}
which leads to the Lagrangian
\begin{eqnarray} \label{eq:1HDMLagNorm}
        \mathscr{L} &=& 
        \mathscr{L}_\mathrm{kin} +
        \mathscr{L}_\mathrm{M} +
        \bigl( D_\mu \phi \bigr)^\dagger\,\bigl( D^\mu \phi \bigr) -
        y_\nu \left( 
         \overline{\ell_\mathrm{L}} \phi \nu_\mathrm{R} + \text{h.c.}
        \right) +
\nonumber\\*
&& {}+
        m^2 \, \phi^\dagger \phi -
        \frac{\lambda}{4} \left( \phi^\dagger \phi \right)^2 \;,
\end{eqnarray}
where we have defined
\begin{equation}
        m^2 := y_\nu^2 \, \widetilde m^2
        \quad \text{and} \quad
        \lambda := y_\nu^4 \, \widetilde\lambda \;.
\end{equation}
Thus, we have recovered the SM extended by right-handed neutrinos,
which proves that the Lagrangian \eqref{eq:4FLag1H} yields exactly the
same low-energy physics, but with additional constraints on the
parameters, namely the compositeness conditions.  From 
eqs.~\eqref{eq:CompositenessUnnorm} and \eqref{eq:HiggsRescaling} we find
that the compositeness conditions for the couplings and the mass
parameter become
\begin{subequations} \label{eq:CompositenessNorm}
\begin{eqnarray}
        \frac{1}{y_\nu^2} &\xrightarrow{\mu\to\Lambda}& 0 \;,
        \label{eq:CompositnessY}
\\
        \frac{m^2}{y_\nu^2} &\xrightarrow{\mu\to\Lambda}& -\frac{1}{G^{(\nu)}}
        \;,
        \label{eq:CompositnessMu1}
\\
        \frac{\lambda}{y_\nu^4} &\xrightarrow{\mu\to\Lambda}& 0 \;.
        \label{eq:CompositnessLambda}
\end{eqnarray}
\end{subequations}
The low-energy neutrino and Higgs masses are now obtained from the RG equations
for the SM extended by right-handed neutrinos. For the RG analysis in see-saw 
models, it is crucial to integrate out the right-handed neutrinos and the 
corresponding part of the neutrino Yukawa coupling matrix at the mass
thresholds. 
The computation of the RG evolution requires the 
$\beta$-functions of all gauge and Yukawa couplings (including those of the neutrinos), 
of the quartic Higgs coupling and of the dimension~5 neutrino mass operator
(see e.g.\ \cite{Chankowski:2001mx,Antusch:2001ck}). 
For completeness, we have listed the relevant \(\beta\)-functions in the
appendix.
When integrating out the right-handed neutrinos, the 
dimension~5 neutrino mass operator has to be matched
at each threshold \cite{Antusch:2002rr}. 
In this framework, we calculate the low-energy parameters,
starting with the compositeness conditions at the condensation scale
and solving the relevant systems of coupled differential equations.
The results for the Higgs mass and for the mass of the neutrino participating
in condensation turn out to
be not very sensitive to the exact boundary conditions
\eqref{eq:CompositnessMu1} and \eqref{eq:CompositnessLambda} due to the
quasi-fixed-point behavior that arises once eq.~\eqref{eq:CompositnessY}
is satisfied.

\subsection{Neutrino Masses at Low Energy} \label{sec:NuMass}
\begin{figure}[ht*]
        \begin{center}
                $\CenterEps[1]{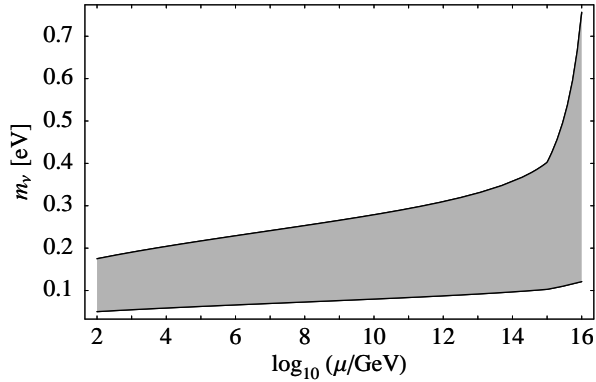}$
        \end{center}
\caption{\label{fig:NeutrinoYukawaCoupling} Running of the mass 
of the neutrino participating in condensation.
The starting values for the neutrino Yukawa coupling are in the range 
$y_\nu \in [2,5]$ at the scale of new physics, $\Lambda$.
The Majorana mass of the right-handed neutrino is 
$M=10^{15}\,\mathrm{GeV}$ 
at $\Lambda$, which we have chosen to be $10^{16}\,\mathrm{GeV}$ in this
example. The gray region shows the possible values of
the neutrino mass with the above 
range of initial values.
The resulting neutrino mass is in agreement with the current limits. 
It can of course be raised or lowered by varying $M$ and further
depends on $\Lambda$, which is a free parameter of the model. 
The kink in the evolution of the neutrino mass corresponds to the
mass threshold at $\mu=M$. Below $M$, the heavy singlet is
integrated out, producing an effective dimension 5
operator which yields a see-saw suppressed Majorana mass. 
}
\end{figure}

The RG evolution of the Majorana mass of the light 
neutrino participating in condensation is shown
in fig.~\ref{fig:NeutrinoYukawaCoupling}.
It can be seen that a large, non-perturbative neutrino Yukawa coupling
is in agreement with the current limits on neutrino masses, if the see-saw
scale $M$ is large enough.  Since the neutrino Yukawa coupling at the
condensation scale has to be non-perturbatively large and 
is thus not a free parameter of the model, 
an allowed range for the neutrino mass translates into an allowed range for 
$M$. 
For instance, with $y_\nu \in [2,5]$, 
$M \in [10^{14}\,\mathrm{GeV},10^{15.5}\,\mathrm{GeV}]$ at the condensation
scale, which we have chosen to be
$\Lambda = 10^{16}\,\mathrm{GeV}$, we obtain a range 
\(m_\nu \in [0.02\,\mathrm{eV},1.36\,\mathrm{eV}]\) 
for the  neutrino mass at low energy. 
Except for the neutrino and the top, the Yukawa couplings
of the fermions have been omitted.

\subsection{Higgs Mass Prediction} \label{sec:HiggsMass}
\begin{figure}[ht*]
        \begin{center}              
                $\CenterEps[1]{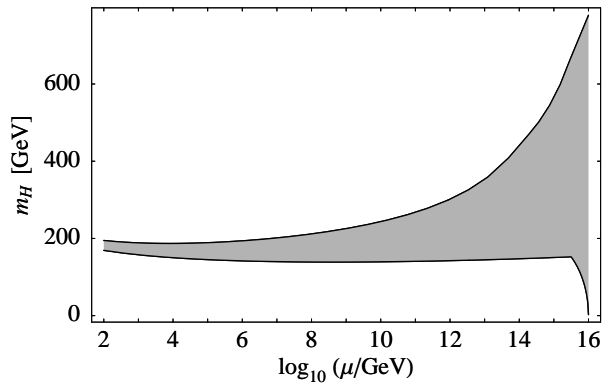}$
        \end{center}
\caption{\label{fig:HiggsMassFP} 
RG evolution of the ``Higgs mass''
\(m_H(\mu)=\sqrt{\lambda(\mu)/2}\,v_\mathrm{EW}\) (with
$v_\mathrm{EW}=246$~GeV),
which equals at the EW scale the physical Higgs mass.  
The quasi-fixed point leads to a rather narrow range for the mass at the
EW scale, in this example 
$170\,\mathrm{GeV} \lesssim m_H \lesssim 195\,\mathrm{GeV}$.
Here we chose $\Lambda=10^{16}\,\mathrm{GeV}$ for the scale of new
physics.  The input parameters at this energy were varied within
the intervals $y_\nu \in [2,5]$, 
$M \in [10^{14}\,\mathrm{GeV},10^{15.5}\,\mathrm{GeV}]$ 
and $\lambda \in [0,20]$.  We further allowed 
$m_t \in [170\,\mathrm{GeV},180\,\mathrm{GeV}]$.
The gray region shows the possible values of
the Higgs mass with these ranges of initial values.
}
\end{figure}

The Higgs mass can be predicted due
to the quasi-fixed point in the RG evolution of the Higgs self-coupling.
The running of the Higgs mass from the condensation scale $\Lambda$
to the EW scale is shown in fig.~\ref{fig:HiggsMassFP} for a wide range
of parameters $M$, $y_\nu$ and $\lambda$ at the scale $\Lambda$. 
For $\Lambda=10^{16}\,\mathrm{GeV}$, we obtain Higgs masses in the range
$170\,\mathrm{GeV} \lesssim m_H \lesssim 195\,\mathrm{GeV}$. 
One should keep
in mind that this prediction depends on the condensation scale $\Lambda$, which
is a free parameter in our model. As in section \ref{sec:NuMass}, 
except for the neutrino and the top, the Yukawa couplings
of the fermions have been omitted.

\clearpage

\section{Three-Neutrino Condensation} \label{sec:3neutrinos}

The model discussed so far can be extended to the case where 
all three generations of neutrinos participate in the condensation.
In general, the low-energy theory contains several
Higgses. However, if the four-fermion couplings satisfy a 
``factorization relation'' \cite{Martin:1991xw},
\begin{equation}
        \sum_{f,g,h,i=1}^3 G^{(\nu)}_{fghi} \, 
        (\overline{ \ell_{\mathrm{L}}^f} \nu_{\mathrm{R}}^g)
        (\overline{ \nu_{\mathrm{R}}^h} \ell_{\mathrm{L}}^i) 
        =
        \biggl( \sum_{f,g=1}^3 h^{(\nu)}_{fg} \,
         \overline{\ell_\mathrm{L}^f} \nu_\mathrm{R}^g \biggr)
        \biggl( \sum_{h,i=1}^3 h^{(\nu)*}_{ih} \, 
         \overline{\nu_\mathrm{R}^h} \ell_\mathrm{L}^i \biggr) \;,
\end{equation}
then by following the steps of 
eqs.~(\ref{eq:4FLag1H}) -- (\ref{eq:hence}),
the Lagrangian can be rewritten in terms of \emph{one} auxiliary field
$\Phi\sim \sum_{fg} h^{(\nu)}_{fg} \,
         \overline{\ell_\mathrm{L}^f} \nu_\mathrm{R}^g$.
Hence we obtain again a one-Higgs model after 
condensation analogous to section \ref{sec:LagUebergang}.  
The gap equations can be treated as in
section~\ref{sec:NeutrinoCondensation}. If three neutrinos
condense, the infrared quasi-fixed point of the RG evolution 
leads to three heavy Dirac mass eigenvalues. 
The Majorana mass term of the right-handed neutrino in eq.~\eqref{eq:MajoranaMass} 
must now be generalized to a mass matrix \(M_{ij}\) for three
right-handed neutrinos, with entries which are unprotected by symmetries, 
leading to three heavy eigenvalues. 
The degeneracy or hierarchy of the see-saw scales is then conveyed
rather directly into the full light neutrino mass pattern.
The Higgs mass is almost unchanged compared to the results of section
\ref{sec:HiggsMass}, in spite of the contribution from the Yukawa
couplings of the additional neutrinos.  Using the same input
parameters as in fig.~\ref{fig:HiggsMassFP}, now with three equally 
large Dirac masses for the neutrinos, we find
$175\,\mathrm{GeV} \lesssim m_H \lesssim 195\,\mathrm{GeV}$.

\section{Combined Neutrino and Top Condensation} \label{sec:2higgs}

Besides the neutrino condensate discussed in section
\ref{sec:NeutrinoCondensation}, there can be a further condensate 
connected to the top quark.  This means that
in addition to the four-fermion coupling of eq.~\eqref{eq:4FNeutrino}
for the neutrino, there is a corresponding term for the top quark,
$
G^{(t)} \, (\overline{ q_{\mathrm{L}}} 
  t_{\mathrm{R}})
  (\overline{ t_{\mathrm{R}}} q_{\mathrm{L}}) 
$,
and a mixed term
$
G^{(\nu t)} \, 
  (\overline{ \ell_{\mathrm{L}}} 
  \nu_{\mathrm{R}})
  (\overline{ t_{\mathrm{R}}} q_{\mathrm{L}})
$.
If $G^{(\nu)},G^{(t)}$ and $G^{(\nu t)}$ become critical, then 
in general two independent condensates form, similar to the case
of combined top and bottom condensation \cite{Luty:1990bg}.
Note that for $(G^{(\nu t)})^2 = G^{(\nu)}\cdot G^{(t)}$ 
this is a one-Higgs scenario, which coincides with the model
discussed by Martin \cite{Martin:1991xw}.
However, in general we are dealing with an effective Two Higgs 
Doublet Model (2HDM), where the effective Lagrangian does not coincide
with the 2HDMs usually discussed, which are phenomenologically 
severely constrained. The scalars which are generated 
dynamically here are a ``leptonic Higgs'', \(\phi_\nu\), and a 
``hadronic Higgs'', \(\phi_t\). The mixed coupling $G^{(\nu t)}$ leads
to a term of the form $ m_{\nu t}^2 \, \phi_\nu^\dagger \phi_t $
in the effective Higgs potential. Besides, four-Higgs interactions with
odd numbers of $\phi_\nu$ and $\phi_t$ become allowed in the low-energy
effective theory.   

\begin{figure}[ht*]
        \begin{center}
       $\CenterEps[1]{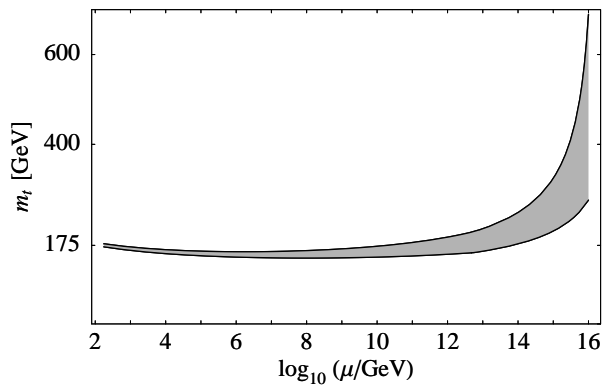}$
        \end{center}
\caption{\label{fig:2HDMTopYukawaCoupling} 
Running of the ``top mass'' \(m_t=y_t(\mu)\,v_\mathrm{EW}\sin(\beta)/\sqrt{2}\)
in a 2HDM with a leptonic and a hadronic Higgs,
formed by a neutrino and a top condensate, and with $\tan \beta = 1.3$
at the electroweak scale.  We have chosen the value of $\tan\beta$
such that the quasi-fixed point of the RG evolution reproduces 
$m_t \approx 175\,\mathrm{GeV}$.  The initial range for
the top Yukawa coupling at the scale of new physics, chosen to be
$\Lambda=10^{16}\,\mathrm{GeV}$ in this example, is $y_t \in [2,5]$.
The gray region contains the resulting values for the top mass.
  }
\end{figure}

An interesting feature of 2HDMs with a leptonic and a hadronic Higgs 
is that the correct top mass can now be obtained dynamically as a 
quasi-fixed point of the RG evolution. This has been found to be 
impossible in minimal models with pure top condensation, where the 
predicted top mass is too large. In the two-Higgs case, the necessary 
additional degree of freedom is the ratio of the vevs, $\tan\beta$. 
A top mass of approximately \(175\,\mathrm{GeV}\) is obtained for 
\(\tan \beta \approx 1\), 
as illustrated in fig.~\ref{fig:2HDMTopYukawaCoupling}.
This also yields large Dirac masses for the neutrinos, but due to the
see-saw suppression their physical masses remain tiny.

Another issue that has to be addressed is the Higgs mass spectrum, as
massless or very light Higgses would be in conflict with experiments.
As suggested by the analysis of the case with a top and a
bottom condensate \cite{Luty:1990bg}, it should be possible to obtain
phenomenologically viable Higgs masses by making the parameter 
$m_{\nu t}^2$ sufficiently large.  Hence, it seems worthwhile to study
2HDMs with neutrino and top condensates in more detail.

\section{Discussion and Conclusions}
\label{sec:conclusion}

In this article we have studied the possibility that the EW
symmetry is broken dynamically by the formation of a neutrino condensate.
We have started from the SM with right-handed neutrinos, but without a fundamental 
Higgs field. Instead of the Higgs field we have postulated strong attractive 
four-fermion interactions. In addition, we have included huge Majorana masses
for the right-handed neutrinos, since there is no protective symmetry.
The analysis of the coupled Dirac-Majorana gap equations has shown that 
non-trivial solutions exist, which dynamically produce a large Dirac neutrino 
mass and which break the EW symmetry in the desired way. However, due to the 
presence of the huge Majorana mass term we obtain physical neutrino masses in 
the phenomenologically allowed region via the see-saw mechanism.
We have illustrated in the auxiliary field formalism that the NJL-type interactions 
lead to a composite scalar sector which resembles a Higgs
Lagrangian with certain boundary conditions, i.e.~predictions. In particular, 
the compositeness conditions require that the Yukawa coupling $y_\nu$ of
the condensing neutrino becomes non-perturbative at high energies.
To evaluate the predictions, we have translated these boundary conditions in 
the framework of the low-energy effective Lagrangian into high-energy boundary
conditions for the renormalization group running. 

For our minimal model discussed in sections~\ref{sec:NeutrinoCondensation} and 
\ref{sec:Phenomenology} the effective Lagrangian is the SM with massive
neutrinos. The boundary conditions and the infrared quasi-fixed-point behavior 
of the Higgs self-coupling and of the relevant Yukawa coupling $y_\nu$ lead to 
two predictions for a given condensation scale $\Lambda$.  For $\Lambda=10^{16}\,\mathrm{GeV}$, 
typical values of $y_\nu(\Lambda)$ and $\lambda(\Lambda)$ consistent with
the boundary conditions, and for
$M \in [10^{14}\,\mathrm{GeV},10^{15.5}\,\mathrm{GeV}]$, 
we have found a Higgs mass in the range 
$170\,\mathrm{GeV} \lesssim m_H \lesssim 195\,\mathrm{GeV}$,
which is in agreement with current experimental bounds.
Moreover, we have found that an upper bound for the neutrino mass 
of the order of $1\,\mathrm{eV}$ ($0.1\,\mathrm{eV}$) translates into a lower bound for the 
see-saw scale $M$ of the order of $10^{14}\,\mathrm{GeV}$
($10^{15}\,\mathrm{GeV}$).

We have also briefly studied the possibility that more than one neutrino
condenses. This leads in general to several Higgses and under a factorization
condition to a one-Higgs scenario. Furthermore, we have outlined 
the possibility of a combined neutrino-top condensation,
which corresponds without further assumptions to a two-Higgs model.
In addition, it might be possible to extend neutrino 
condensation to a supersymmetric model.

We have not attempted to embed these scenarios into a larger framework where
the four-fermion terms are generated in strongly coupled broken gauge 
theories as it was done in ``top color'' theories \cite{Hill:1991at}
in the case of top condensation.\footnote{%
Note that such scenarios can possess the attractive property of
generating Yukawa couplings via gauge couplings.}
This should be interesting, since such a 
framework would, for example, allow to address the question if the 
gauge couplings of the extended gauge sector unify above the 
condensation scale. In addition, threshold effects near the condensation scale 
(in combination with extended gauge sectors or by themselves) might affect 
unification. Such threshold corrections are generally expected to 
be large in this non-perturbative scenario.

To conclude, we have introduced a dynamical realization of electroweak
symmetry breaking with massive neutrinos, where the SM Higgs particle
emerges from neutrino condensation, leading to predictions for the 
Higgs mass and for the see-saw scale.

\section*{Acknowledgements}
This work was supported in part by the 
``Sonderforschungsbereich~375 f\"ur Astro-Teilchenphysik der 
Deutschen Forschungsgemeinschaft''.

\section*{Appendix: Renormalization Group Equations}

In the SM extended by right-handed neutrinos
one has to consider several effective theories,
corresponding to the ranges between the non-degenerate eigenvalues of
the Majorana mass matrix \(M\).
At the thresholds,
the heavy degrees of freedom are successively integrated out.
As introduced in \cite{Antusch:2002rr}, a superscript \((n)\) denotes 
a quantity between the $n$th and the \((n+1)\)th mass threshold.
With this definition, the \(\beta\)-functions for 
the Yukawa coupling matrices are given by
 \begin{subequations}
 \begin{eqnarray}
        16\pi^2 \accentset{(n)}{\beta}_{Y_\nu}
        &=&
        \accentset{(n)}{Y}_\nu \left\{
        \frac{3}{2} \RaiseBrace{\bigl(}
        \accentset{(n)}{Y}^\dagger_\nu\accentset{(n)}{Y}_\nu\RaiseBrace{\bigr)}
        - \frac{3}{2}(Y_e^\dagger Y_e)
        -\frac{3}{4} g_1^2 -\frac{9}{4} g_2^2
         \right.
        \nonumber\\*
        &&\hphantom{\accentset{(n)}{Y}_\nu \left[ \right.} \left.
        {}+\Tr\left[Y_e^\dagger Y_e
                +\accentset{(n)}{Y}_\nu^\dagger \accentset{(n)}{Y}_\nu
                +3\,Y_d^\dagger Y_d\
                +3\,Y_u^\dagger Y_u\right]
        \right\}
        \;,\\
        16\pi^2 \, \accentset{(n)}{\beta}_{Y_e}
        & = &
        Y_e
        \left\{ 
        \vphantom{\Tr\left[Y_e^\dagger Y_e
                +z_\nu^{(1)}\,\accentset{(n)}{Y}_\nu^\dagger \accentset{(n)}{Y}_\nu
                +3z_d^{(1)}\,Y_d^\dagger Y_d\
                +3z_u^{(1)}\,Y_u^\dagger Y_u\right]}
     \frac{3}{2} Y_e^\dagger Y_e 
         -\frac{3}{2}\, \accentset{(n)}{Y}_\nu^\dagger \accentset{(n)}{Y}_\nu 
         -       \frac{15}{4} g_1^2 - \frac{9}{4} g_2^2
         \right.\nonumber\\*
        & &\hphantom{Y_e\left\{ \right.}
         \left.{}+
         \Tr\left[Y_e^\dagger Y_e
                +\accentset{(n)}{Y}_\nu^\dagger \accentset{(n)}{Y}_\nu
                +3\,Y_d^\dagger Y_d\
                +3\,Y_u^\dagger Y_u\right]
        \right\}\;,
        \\
        16\pi^2 \, \accentset{(n)}{\beta}_{Y_d}
        & = &
        Y_d
        \left\{ 
     \frac{3}{2} Y_d^\dagger Y_d 
         -\frac{3}{2}\, Y_u^\dagger Y_u          
         - \frac{5}{12} g_1^2 - \frac{9}{4} g_2^2 - 8\,g_3^2
         \right.\nonumber\\*
        & &\hphantom{Y_e\left\{ \right.}
         \left.{}+ \Tr\left[Y_e^\dagger Y_e
                +\accentset{(n)}{Y}_\nu^\dagger \accentset{(n)}{Y}_\nu
                +3\,Y_d^\dagger Y_d
                +3\,Y_u^\dagger Y_u\right]
        \right\}\;, 
        \\
        16\pi^2 \, \accentset{(n)}{\beta}_{Y_u}
        & = &
        Y_u
        \left\{ 
     \frac{3}{2} Y_u^\dagger Y_u 
         - \frac{3}{2}\, Y_d^\dagger Y_d  
         - \frac{17}{12} g_1^2 - \frac{9}{4} g_2^2 - 8\,g_3^2
         \right.\nonumber\\*
        & &\hphantom{Y_e\left\{ \right.}
         \left.{}+ \Tr\left[Y_e^\dagger Y_e
                +\accentset{(n)}{Y}_\nu^\dagger \accentset{(n)}{Y}_\nu
                +3\,Y_d^\dagger Y_d
                +3\,Y_u^\dagger Y_u\right]
        \right\}\;.
 \end{eqnarray}
 \end{subequations}
The \(\beta\)-function for the Majorana mass matrix \(M\) reads
\begin{eqnarray}\label{eq:RGEForMBetweenThresholds}
16\pi^2 \accentset{(n)}{\beta}_{M} = \RaiseBrace{\bigl(}\accentset{(n)}{Y}_\nu   
   \accentset{(n)}{Y}^\dagger_\nu \RaiseBrace{\bigr)}\, \accentset{(n)}{M} 
   + \accentset{(n)}{M}\,\RaiseBrace{\bigl(}\accentset{(n)}{Y}_\nu   
   \accentset{(n)}{Y}^\dagger_\nu \RaiseBrace{\bigr)}^T \;,
\end{eqnarray}
and the RG evolution of the quartic Higgs self-coupling is
determined by\footnote{To our knowledge, this \(\beta\)-function
has not yet been written explicitly in the literature.}
\begin{eqnarray}
\lefteqn{ 16\pi^2\,\accentset{(n)}{\beta}_{\lambda}
 = 
 6\,\lambda^2 
 -3\,\lambda\,(3g_2^2+g_1^2)
 +3\,g_2^4
 +\frac{3}{2}\,(g_1^2+g_2^2)^2
 }\nonumber
 \\*
 & &{}
 +4\,\lambda\,
 \Tr\left[
       Y_e^\dagger Y_e
       +\accentset{(n)}{Y}_\nu^\dagger\accentset{(n)}{Y}_\nu
       +3\,Y_d^\dagger Y_d
       +3\,Y_u^\dagger Y_u
 \right]
 \\*
 & &{}
 -8\,\Tr\left[
  Y_e^\dagger Y_e\,Y_e^\dagger Y_e
  +    \accentset{(n)}{Y}_\nu^\dagger\accentset{(n)}{Y}_\nu
               \,\accentset{(n)}{Y}_\nu^\dagger\accentset{(n)}{Y}_\nu
       +3\,Y_d^\dagger Y_d\,Y_d^\dagger Y_d
       +3\,Y_u^\dagger Y_u\,Y_u^\dagger Y_u
 \right]\;.\nonumber
\end{eqnarray}
Finally, the \(\beta\)-function for the 
effective neutrino mass operator reads
\begin{eqnarray}\label{eq:RGEForKappaBetweenThresholds}
16\pi^2\accentset{(n)}{\beta}_\kappa & = & 
 -\frac{3}{2} (Y_e^\dagger Y_e)^T \:\accentset{(n)}{\kappa}
 -\frac{3}{2}\,\accentset{(n)}{\kappa} \, (Y_e^\dagger Y_e)
 + \frac{1}{2} \RaiseBrace{\bigl(} \accentset{(n)}{Y}^\dagger_\nu   
   \accentset{(n)}{Y}_\nu \RaiseBrace{\bigr)}^T \,
  \accentset{(n)}{\kappa}
 +\frac{1}{2}\,\accentset{(n)}{\kappa} \: \RaiseBrace{\bigl(}
 \accentset{(n)}{Y}^\dagger_\nu\accentset{(n)}{Y}_\nu\RaiseBrace{\bigr)}
\nonumber \\*
&& {}
 +2\,\Tr(Y_e^\dagger Y_e)\,\accentset{(n)}{\kappa} 
 +2\, \Tr \RaiseBrace{\bigl(} \accentset{(n)}{Y}^{\dagger}_\nu 
 \accentset{(n)}{Y}_\nu\RaiseBrace{\bigr)}\,\accentset{(n)}{\kappa} 
 +6\,\Tr(Y_u^\dagger Y_u)\,\accentset{(n)}{\kappa} 
  \nonumber \\*
 && {} 
 +6\,\Tr(Y_d^\dagger Y_d)\,\accentset{(n)}{\kappa}
- 3 g_2^2\: \accentset{(n)}{\kappa}
 +\lambda\accentset{(n)}{\kappa}
 \;.
\end{eqnarray}
The one-loop $\beta$-functions for the gauge couplings are of course unchanged
compared to the SM.

\providecommand{\bysame}{\leavevmode\hbox to3em{\hrulefill}\thinspace}

\end{document}